\documentclass[twoside]{article}

\usepackage{Proc_NTSE}
\usepackage{amsthm}
\usepackage{amsfonts}
\usepackage{mathrsfs}
\usepackage{bm}
\usepackage{xcolor}
\usepackage{graphicx}
\usepackage{subfig} 	
\usepackage{relsize}	
\usepackage{array}
\usepackage{float}
\usepackage{color}
\usepackage{booktabs}
\usepackage{cite} 

\def\dd{{\mathrm{d}}}
\def\imag{{\mathrm{i}}}

\mathchardef\-="2D

\def\K{{\mathcal{K}}}
\def\N{{\mathcal{N}_{\max}}}

\def\J{{\mathcal{J}}}
\newcommand{\comments}[1] {}
\newcommand{\half}[1][1] {\mathsmaller{\frac{#1}{2}}}
\newcommand{\bra}[1] {\left< #1 |\right.}
\newcommand{\ket}[1] {\left.| #1 \right> }

\makeatletter

\newcommand{\Rmnum}[1]{\expandafter\@slowromancap\romannumeral #1@}
\makeatother

\begin{document}
\thispagestyle{plain}

\begin{center}
{\Large \bf \strut
 Introduction to Basis Light-Front Quantization Approach to QCD Bound State Problems
\strut}\\
\vspace{10mm}
{\large \bf
Yang Li, Paul W. Wiecki, Xingbo Zhao, Pieter Maris and James P. Vary
}
\end{center}

\noindent{
\small \it Department of Physics and Astronomy, Iowa State University, Ames, IA, 50011}

\markboth{
Yang Li, Paul W. Wiecki, Xingbo Zhao, Pieter Maris and James P. Vary}
{
Introduction to Basis Light-Front Quantization}

\begin{abstract}

Basis Light-Front Quantized Field Theory (BLFQ) is an \textit{ab intio} Hamiltonian approach that adopts light-cone gauge, light-front
quantization and state-of-the-art many-body methods to solve non-perturbative quantum field theory problems. By a suitable choice of
basis, BLFQ retains the underlying symmetries to the extent allowed within light-front coordinates. In this talk, we outline the
scheme for applying BLFQ to QCD bound state problems. We adopt a 2D Harmonic Oscillator with 1D plane wave basis that corresponds to the
AdS/QCD soft-wall solution. Exact treatment of the symmetries will be discussed.
\\[\baselineskip]
{\bf Keywords:} {\it Light-Front; Harmonic Oscillator Basis; QCD; Non-Perturbative; Symmetry}
\end{abstract}

\section{Introduction}

Solving bound state problems arising in quantum chromodynamics (QCD) is the key to understand a series of important questions in physics.
The solutions will provide consistent descriptions of the structure of mesons, baryons and also particles with ``exotic'' quanta
beyond the scope of the constituent quark model. One salient challenge is to predict the spin content of the baryons. Furthermore, it
could also help to explain the nature of confinement and dynamical chiral symmetry breaking. QCD bound states are strong coupling
non-perturbative solutions that cannot be generated from perturbation theory. Among various non-perturbative methods, light-front
Hamiltonian quantization within a basis function approach has shown significant promise by capitalizing on both the advantages of
light-front dynamics as well as the recent theoretical and computational achievements in quantum many-body theory. We begin with an
overview of the light-front quantum field theory. We will then introduce the Basis Light-Front Quantized Field theory (BLFQ) and its
application to bound state problems in quantum field theory.

\section{Light-front quantum field theory}

The idea of quantization on a light-front surface was first considered by Paul Dirac in 1949 in his famous investigation of forms of
relativistic dynamics \cite{Dirac49}. Light-front quantum field theory is quantized on a light-front plane $x^+ \equiv x^0+x^3 = 0$
and evolves according to light-front time $x^+$. It is convenient to define light-front variables $x^\pm = x^0 \pm x^3, \bm{x}^\perp
 = (x^1, x^2)$, where $x^+$ is the light-front time, $x^-$ is the longitudinal coordinate. The light-front momentum $p^\pm = p^0 \pm p^3,
\bm{p}^\perp = (p^1, p^2)$, where the $p^+$ is the longitudinal momentum and $p_+ = \half p^-$ is the light-front energy. For positive
energy states, $p^+$ and $p^-$ are positive. An important consequence of this is that the light-front vacuum state is trivial
\cite{Perry90, Brodsky98}.

Let $M^{\mu\nu}, P^\mu$ be the generators of the Poincar\'e symmetry. $J^k = \half \epsilon^{ijk}M^{ij}, K^i = M^{0i},
(i,j,k = 1,2,3)$ are the equal-time angular momentum and boost generators respectively. The light-front rotation and boost generators
 are $F^- \equiv M^{12} = J^3$, ${F^i} \equiv \varepsilon^{ij}{M}^{j-} = J^i + \varepsilon^{ij}K^j, (i,j=1,2)$ and $E^- \equiv \half
 M^{+-} = K^3$, $E^i \equiv M^{+i} = K^i + \varepsilon^{ij}J^j, (i,j=1,2) $. According to Dirac, in light-front dynamics the number
of kinematic operators of the Poincar\'e algebra  is maximal: $\{ P^+, \bm{P}^\perp, E^-, \bm{E}^\perp, F^- \}$. 
The kinematic feature of
the the light-front boost generators $E^-, \bm{E}^\perp$ provides convenience in evaluating matrix elements of certain experimental
observables where the initial and final states differ by a boost. Note however that the total angular momentum operator is dynamic
in light-front dynamics.

Irreducible representations can be identified with mutually commuting operators or compatible operators. It is customary to take the
set of compatible operators as $\{P^2, W^2, P^+, \bm{P}^\perp, \J^3\}$, where $W^\mu = -\frac{1}{2}\varepsilon^{\mu\nu\kappa\lambda}
M_{\nu\kappa}P_\lambda$ is the Pauli-Lubanski vector, $\J^3 \equiv \frac{W^+}{P^+} = J^3 + \varepsilon^{ij} \frac{E^i P^j}{P^+}$ is the
longitudinal projection of the light-front spin \cite{Brodsky98}. Note that in relativistic dynamics, the total angular momentum
operator $\bm{J}^2 = J_1^2 + J_2^2 + J_3^2$ is generally different from the total spin operator\footnotemark[1] $\bm{\J}^2 = \J_1^2 + \J_2^2 + \J_3^2
 = -W^2/P^2$ \cite{Brodsky98, Bogolubov75}. In light-front dynamics, $P^2 \equiv P_\mu P^\mu, W^2 \equiv W_\mu W^\mu$ are dynamical. 
They have to be diagonalized
 simultaneously at $x^+ = 0$:
\begin{subequations}\label{soe}
 \begin{align}
  P^2 \left.|\mathscr{M}, \J\right> &= \mathscr{M}^2 \left.|\mathscr{M}, \J\right> , \label{soe_a} \\
  W^2 \left.|\mathscr{M}, \J\right> &= -\mathscr{M}^2 \J(\J+1) \left.|\mathscr{M}, \J\right> \label{soe_b}.
 \end{align}
\end{subequations}
\footnotetext[1]{Note that the total spin $\J$ is the observable quantity quoted as ``$J$'' in the Particle Data Group  compilations \cite{pdg}.}
It is conventional to call $P^2$ the ``light-cone Hamiltonian'' $H_\textsc{lc} \equiv P^2$. It is convenient to express this
light-cone Hamiltonian in terms of kinetic energy and potential energy, $H_\textsc{lc} = H_\textsc{lc}^{(0)} + V_{int}$. The kinetic
energy, $H^{(0)}_\textsc{lc}$, resembles the non-relativistic kinetic energy,
  \begin{equation}
 \begin{split}
  H^{(0)}_\textsc{lc} &= 2P^+ P^{(0)}_+ - {\bm{P}^\perp}^2  = \sum_a \frac{{\bm{p}_a^\perp}^2+m^2_a}{x_a} - {\bm{P}^\perp}^2,
 \end{split}
 \end{equation}
where ``$a$'' represents the quark or gluon constituent and $\bm{P}^\perp = \sum_a \bm{p}_a^\perp$ is the total transverse
Center-of-Mass (CM) momentum while $x_a = \frac{p^+_a}{P^+}$ is the longitudinal momentum fraction carried by each constituent.

The triviality of the Fock space vacuum provides a strong appeal for the Fock space representation of the quantum field theory
\cite{Perry90}. The Fock space expansion of an eigenstate in a plane-wave basis reads,
\begin{multline}
 \ket{\varPsi; P} = \sum_{n=0}^\infty
 \sum_{\sigma_1,\cdots\sigma_n}\int\frac{\dd^3{k}_1}{(2\pi)^3 2k^+_1}\theta(k^+_1)
\cdots\frac{\dd^3{k}_n}{(2\pi)^3 2k^+_n}\theta(k^+_n) \\
  2P^+ (2\pi)^3 \delta^3(k_1+\cdots k_n - P)\varPsi_n^{\sigma_1,\cdots\sigma_n}(k_1,\cdots k_n)
a^\dagger_{\sigma_1}(k_1)\cdots a^\dagger_{\sigma_n}(k_n)\ket{0},
\end{multline}
where $\mathrm{d}^3 k = \mathrm{d}^2k^\perp \mathrm{d} k^+$, $\delta^3(p_1-p_2) = \delta(p_1^+-p_2^+)\delta^2(\bm{p}_1^\perp -
\bm{p}_2^\perp)$ and $\varPsi_n^{\sigma_1,\cdots\sigma_n}(k_1,k_2,\cdots k_n)$ is called the light-front wavefunction (LFWF).
LFWFs are boost-invariant, namely frame independent, following the boost-invariance of the light-cone Hamiltonian and the pure
kinematic character of the light-front boosts. In practice, a non-perturbative diagonalization of the Hamiltonian can only
be achieved in a finite-truncated Fock space. The Tamm-Dancoff approximation (TDA) introduces a truncation based on Fock sectors.
The rationale of the TDA is founded on the success of the constituent quark model, according to which the hadrons can be approximated
by a few particles as in the leading Fock space representation \cite{Wilson94}. However, there are also new challenges in
light-front TDA. Fock sector truncation breaks rotational symmetry. As a result, the total spin $\J$ in a truncated
calculation is no longer a good quantum number.

\section{Basis light-front quantized field theory}

Basis Light-Front Quantization (BLFQ) is an \textit{ab inito} Hamiltonian approach to light-front quantum field theory
that adopts a complete set of orthonormal single-particle basis functions for field expansions resulting in Fock space basis states
$\ket{\phi_i}$ expressed in terms of these single-particle basis states \cite{Vary10}. In the Fock space basis, the Hamiltonian and
eigenstates become,
\begin{equation}
 H_{ij} = \bra{\phi_i} H_\textsc{lc}\ket{\phi_j} , \quad \ket{\varPsi} = \sum_i c_i \ket{\phi_i}.
\end{equation}
The system of equations (\ref{soe_a}) is reformulated as a standard eigenvalue problem,
\begin{equation}\label{eigenvalue}
 \sum_j H_{ij} c_j = \lambda c_i,
\end{equation}
which is then truncated and solved numerically. The full field theory is restored in the continuum limit and the complete Fock sectors
limit of the Hamiltonian many-body dynamics.

In principle, the choice of the basis functions is arbitrary but subject to the conditions of completeness and orthonormality. Basis
functions preserving the kinematic symmetries can dramatically reduce the dimensionality of the problem for a specific accuracy. Basis
functions emulating the correct asymptotic behavior of the solution can accelerate the convergence. BLFQ adopts a light-front basis
comprised of plane-wave functions in the longitudinal direction and 2D Harmonic Oscillator (HO) functions in the transverse direction.
The transverse HO basis states are generated by the following operator \cite{Maris13},
\begin{equation}
 P_+^\textsc{ho} = \sum_a \frac{{\bm{p}^\perp_a}^2}{2p^+_a} + \frac{\Omega^2}{2}p_a^+ {\bm{r}^\perp_a}^2,
\end{equation}
where $\bm{r}^\perp_a \equiv \bm{x}_a^\perp = -\frac{\bm{E}^\perp_a}{p^+_a} $ (at $x^+ = 0$) is the transverse
position operator\footnote[2]{Recall the transverse light-front boost at $x^+ = 0$: $E^i = M^{+i} = x^+ P^i - x^i P^+ = -x^i P^+$.}. 
The adoption of the BLFQ basis exploits known similarity between light-front dynamics and non-relativistic dynamics, and is consistent with
the recent development of the AdS/QCD \cite{Brodsky06, Brodsky08}. 
In momentum space, the single-particle basis functions are given in terms of the generalized Laguerre polynomials $L_n^\alpha$ by,
\begin{equation}
 \begin{split}
 \left< p^+,\bm{p}^\perp | n,m,x \right> & = \mathscr{N} e^{\imag m \theta}  \left(\mathsmaller{\frac{\rho}{\sqrt{x}}}\right)^{|m|}
e^{-\frac{\rho^2}{2x}} L_n^{|m|}(\rho^2/x) \delta(p^+ - x P^+)  \\
                                         & \equiv \frac{1}{\sqrt{x}}\Psi_n^m( \mathsmaller{\frac{\bm{p}^\perp}{\sqrt{x}}} ) 2\pi 2p^+
\delta(p^+ - x P^+) , \quad \rho=\frac{|\bm{p}^\perp|}{\sqrt{P^+\Omega}}, \theta=\arg\bm{p}^\perp,
 \end{split}
\end{equation}
which are associated with the HO eigenvalues $E_{n,m} = ( 2n + |m| + 1) \Omega$. $x$ is the longitudinal momentum fraction.
We can identify the HO energy scale parameter $b = \sqrt{P^+\Omega}$ comparing with $b = \sqrt{M\Omega}$ used in the non-relativistic
HO basis \cite{Honkanen11, Zhao12, Wiecki13}. The orthonormality of the basis functions reads,
\begin{equation}
\begin{split}
 \left< n, m, x | n', m', x' \right>
 &= 2\pi 2x \delta(x-x') \delta_{n,n'} \delta_{m,m'}.
\end{split}
\end{equation}

To introduce a finite truncation, BLFQ selects a particular finite subset of the basis space. In the longitudinal direction, we
confine the longitudinal coordinate $-\mathsf{L} \le x^- \le +\mathsf{L}$ with \textit{periodic boundary condition} for bosons and \textit{anti-periodic
boundary condition} for fermions. Thus the longitudinal momentum $p^+$ is discretized as,
\begin{equation}
p^+ = \frac{2\pi k}{\mathsf{L}}, \quad k = \left\{
 \begin{array}{ll}
 0, 1, 2, 3, \cdots & \quad \text{for bosons} \\
 \frac{1}{2}, \frac{3}{2}, \frac{5}{2}, \frac{7}{2}, \cdots & \quad \text{for fermions}, \\
 \end{array}
\right.
\end{equation}
where $\mathsf{L}$ is the length of the longitudinal box. We omit the zero-modes for the bosons in the calculations that follow. In the transverse
direction, we select the Fock space basis states by,
\begin{equation}
  \sum_a \left( 2n_a + |m_a| + 1 \right) \le \mathcal{N}_{\max}.
\end{equation}
Let $\mathcal{P}$ denote the projection operator for the truncated basis space. Then for a basis state $\ket{\alpha} \equiv
\ket{ n_1,m_1,x_1, n_2,m_2,x_2,\cdots}$, $ \mathcal{P} \ket{\alpha} = \theta( \N-N_\alpha)\ket{\alpha}$ where $N_\alpha
= \sum_a (2n_a+|m_a|+1)$. The continuum limit is achieved by taking $\mathsf{L} \to \infty, \N \to \infty$.

A symmetry may be broken by the basis truncation. For example, let $A$ be a conserved operator and $[H_\textsc{lc}, A] = 0$. Then
in the truncated basis space, the commutator becomes \[ [\mathcal{P}H_\textsc{lc}\mathcal{P}, \mathcal{P}A\mathcal{P}] = \mathcal{P}
\left[ [\mathcal{P}, H_\textsc{lc}], [\mathcal{P},A]\right]\mathcal{P}. \] For the transverse truncation, the commutation relation survives
if $[A, P_+^\textsc{ho}] = 0$. The proof is as following: $[A, P_+^\textsc{ho}] = 0 \;
\Rightarrow \; A \ket{\alpha} = \sum_{\alpha'} C_{\alpha'} \delta_{N_\alpha, N_{\alpha'}} \ket{\alpha'}$. Then
 \begin{eqnarray*}
 \label{truncation}
\mathcal{P} A \ket{\alpha} &=& \mathcal{P} \sum_{\alpha'} C_{\alpha'} \delta_{N_\alpha,N_{\alpha'}}\ket{\alpha'}
=  \sum_{\alpha'} C_{\alpha'} \delta_{N_\alpha,N_{\alpha'}} \theta( \N -N_{\alpha'}) \ket{\alpha'} \\
&=&  \theta( \N - N_{\alpha}) \sum_{\alpha'} C_{\alpha'} \delta_{N_\alpha,N_{\alpha'}}  \ket{\alpha'}
=  A\mathcal{P} \ket{\alpha} \\
\quad \Rightarrow &\;&  [\mathcal{P}, A] = 0 \\
\quad \Rightarrow &\;&  [\mathcal{P}H_\textsc{lc} \mathcal{P}, \mathcal{P}A\mathcal{P}] = 0.
\end{eqnarray*}

Among all generators of the Poincar\'e symmetry, a complete set of compatible operators (including the Hamiltonian) is particularly useful for
solving Eq.~(\ref{eigenvalue}), as the Hamiltonian is block diagonal with respect to the mutual eigenstates. A distinguishing feature
of the BLFQ basis with the finite truncation is that it preserves the set of compatible kinematic operators. We define
$P_+^\textsc{cm} = \frac{1}{2P^+}({\bm{P}^\perp}^2 + \Omega^2 {\bm{E}^\perp}^2) = \frac{1}{2 P^+}{\bm{P}^\perp}^2 +  \frac{\Omega^2}{2}P^+
{\bm{R}^\perp}^2 $. With some effort, one can show that $\{H_\textsc{lc}, P^+, P_+^\textsc{cm}, J^3, \J^3 \}$ is a set of mutually commuting
operators and $[P^+, P_+^\textsc{ho}] = [J^3, P_+^\textsc{ho}] = [P_+^\textsc{cm}, P_+^\textsc{ho}] = [\J^3, P_+^\textsc{ho}] = 0$,
where $\J^3 = J^3 + {\frac{\varepsilon^{ij}}{P^+}} E^iP^j = J^3 - \varepsilon^{ij}R^iP^j$. Therefore in the finite basis space the truncated
operators,
 \begin{equation}
\{ \mathcal{P}H_\textsc{lc}\mathcal{P} , \mathcal{P}P^+\mathcal{P}, \mathcal{P}P_+^\textsc{cm}\mathcal{P}, \mathcal{P}J^3\mathcal{P},
\mathcal{P}\J^3\mathcal{P}\},
 \end{equation}
 still form a set of compatible operators.

For simplicity, we will omit the projection operator when there is no danger of confusion. For example, the compatible operators
in the finite-truncated basis space will be denoted as, $\{ H_\textsc{lc} , P^+, P_+^\textsc{cm}, J^3, \J^3\}$. The compatible
operators allow us to further fix the total longitudinal momentum and the angular momentum projection of the system from the
kinematic construction:
\begin{equation}
 \begin{split}
 P^+ &= \frac{2\pi \mathcal{K}}{\mathsf{L}} \quad \Rightarrow \quad \sum_a k_a = \mathcal{K};\\
 J^3 &= M_j \quad \Rightarrow  \quad \sum_a \left( m_a + \sigma_a \right) = M_j,
\end{split}
\end{equation}
where $\sigma_a$ is the spin projection of the $a$-th constituent.
Due to the boost invariance, the light-cone Hamiltonian depends on the longitudinal momentum fractions $x_a = \frac{k_a}{\K}$ instead of
$p_a^+$. Therefore, for a fixed $\mathsf{L}$, the continuum limit is also achieved by taking $\K \to \infty, \N \to \infty$.

We also take advantage of the internal symmetries to fix the charge $Q = \sum_a Q_a$, baryon number $B = \sum_a B_a$, color projection
and the total color, that is, color configurations are restricted to color-singlet configurations.

Let HO states $\ket{N,M}$ be the mutual eigenstates of $P_+^\textsc{cm}$ and $L^3 \equiv J^3 - \J^3 = R^1P^2 - R^2P^1$ associated with the
eigenvalues $E_\textsc{cm} = E_{N,M} \equiv (2N+|M|+1)\Omega$ and $M$, respectively. Then the eigenvalue $\mathcal{M}_j$ of the light-front 
spin projection $\J^3$ can be expressed in terms of the eigenvalues of $J^3$ and $L^3$ as $\mathcal{M}_j = M_j - M$.
The eigenstates of $P^2$ can be identified with the eigenvalues of the compatible operators as $\ket{\mathscr{M}^2, \frac{2\pi\K}{L},
E_{N,M}, M_j, \mathcal{M}_j}$. They can also be identified in terms of $N,M$ as
$\ket{\mathscr{M}, \K, \mathcal{M}_j, N, M} \equiv \ket{\varphi_\text{intr}} \otimes \ket{\Phi_\textsc{cm}}$, where $\ket{\varphi_\text{intr}} =
\ket{\mathscr{M}, \K, \mathcal{M}_j}, \ket{\Phi_\textsc{cm}} = \ket{N,M}$. Therefore, the finite BLFQ basis admits an exact factorization
of the center-of-mass motion \cite{Caprio12}.

\begin{figure}[h]
\centering
 \includegraphics[width=0.9\textwidth]{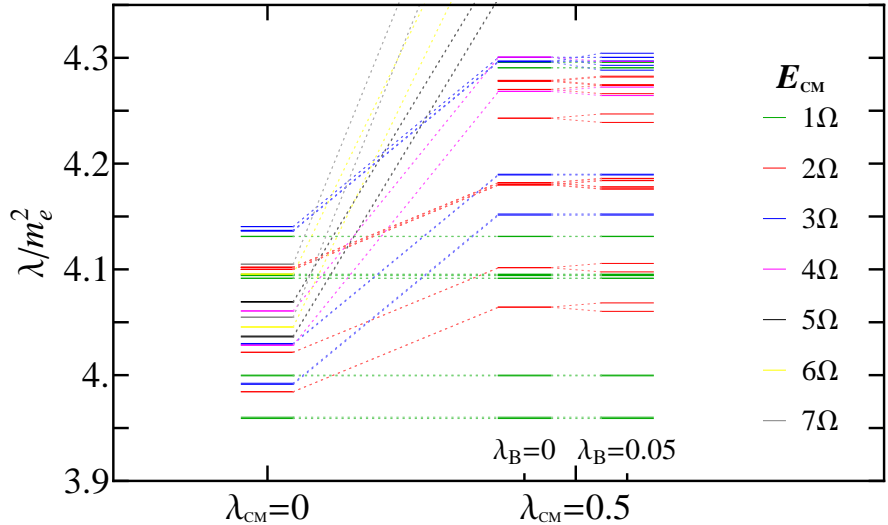}
\caption{(Color Online) Using Lagrange multiplier method to lift the CM excitations. The calculation is performed in the Fock sectors
 $\ket{e^+ e^-}$ and $\ket{e^+e^-\gamma}$ with $M_j = 0, \K = \N = 8, \sqrt{P^+\Omega} = 0.3 m_e$ (See Section 4 for details). The
vertical axis shows the eigenvalues of the light-cone Hamiltonian $H_\textsc{lc}$. States with different $E_\textsc{cm}$ are colored
differently. We show spectra with the lowest 50 states for three cases: $\lambda_\textsc{cm} = 0,\lambda_\textsc{B} = 0$; $\lambda_\textsc{cm} = 0.5,
\lambda_\textsc{b} = 0$ and  $\lambda_\textsc{cm} = 0.5, \lambda_\textsc{b} = 0.05$. In the last case, degeneracies caused by CM
excitations are lifted.
}
\label{lagrange multiplier}
\end{figure}

The light-cone Hamiltonian $H_\textsc{lc} = H^{(0)}_\textsc{lc} + V_{int} $ only acts on the intrinsic part of the wavefunction.
So different CM states are degenerate. It is useful to introduce Lagrange multipliers in the Hamiltonian \cite{Gloeckner74}:
\begin{equation}
P_+ \to P_+ + \lambda_\textsc{cm} (P_+^\textsc{cm} -  \Omega) + \lambda_\textsc{b} \Omega L^3,
\end{equation}
where $\lambda_\textsc{cm} > 0$ and $|\lambda_\textsc{b}| < \lambda_\textsc{cm}$. The eigenvalues of the light-cone
Hamiltonian become $\lambda = \mathscr{M}^2 + 2\lambda_\textsc{cm} (2 N + |M|)P^+\Omega + 2\lambda_\textsc{b} M P^+\Omega$. 
The CM excitations are driven to high energy with sufficiently large $\lambda_\textsc{cm}$. Fig. \ref{lagrange multiplier} shows
the use of Lagrange multipliers to lift the degeneracies of the states in a numerical calculation of a positronium model. 

To separate the CM part and the intrinsic part in the basis functions in single-particle coordinates, we introduce the generalized 2D
Talmi-Moshinsky (TM) transformation \cite{Talmi-Moshinsky},
\begin{multline}
 \Psi_{n_1}^{m_1}(\mathsmaller{\frac{\bm{p}_1}{\sqrt{x_1}}}) \Psi_{n_2}^{m_2}(\mathsmaller{\frac{\bm{p}_2}{\sqrt{x_2}}}) = \\
 \sum_{NMnm} \left.\left({NMnm}|{n_1m_1n_2m_2}\right)\right|_{\xi}
\delta_{\varepsilon_1+\varepsilon_2, E+\epsilon} \delta_{M+m, m_1+m_2}
\Psi_N^M(\mathsmaller{\frac{\bm{p}_1+\bm{p}_2}{\sqrt{x_1+x_2}}})\Psi_n^m(\mathsmaller{\frac{\bm{p}}{\sqrt{x}}}),
\end{multline}
where $\varepsilon_i = 2n_i+|m_i|+1, E = 2N+|M|+1, \epsilon = 2n+|m|+1, \xi = \arctan \sqrt{\frac{x_2}{x_1}},
x = \mathsmaller{\sqrt{\frac{x_1x_2}{x_1+x_2}}}$ and $\bm{p} = \frac{x_2}{x_1+x_2}\bm{p}_1 - \frac{x_1}{x_1+x_2}\bm{p}_2$
is the intrinsic momentum. The sum is finite due to the Kronecker delta $\delta_{E_1+E_2, E+\varepsilon}$. The coefficients
$\left.\left({NMnm}|{n_1m_1n_2m_2}\right)\right|_{\xi}$ are called the generalized 2D TM coefficients. They can be computed
analytically within a closed form (See appendix A).

\section{Basis light-front quantization applied to QED}

In recent applications to QED, Honkanen et al. and Zhao, et al. calculated the electron anomalous magnetic moment and obtained high precision results
\cite{Honkanen11, Zhao12}.
Here we present an application of BLFQ to a highly regularized model of positronium in the Fock sectors $\ket{e^- e^+}$ and $\ket{e^- e^+\gamma}$
which complements the treatment of positronium presented in Ref. \cite{Wiecki13}.

\begin{figure}
\centering
 \subfloat[$e \to e \gamma$ vertex ]{\includegraphics[width=0.25\textwidth]{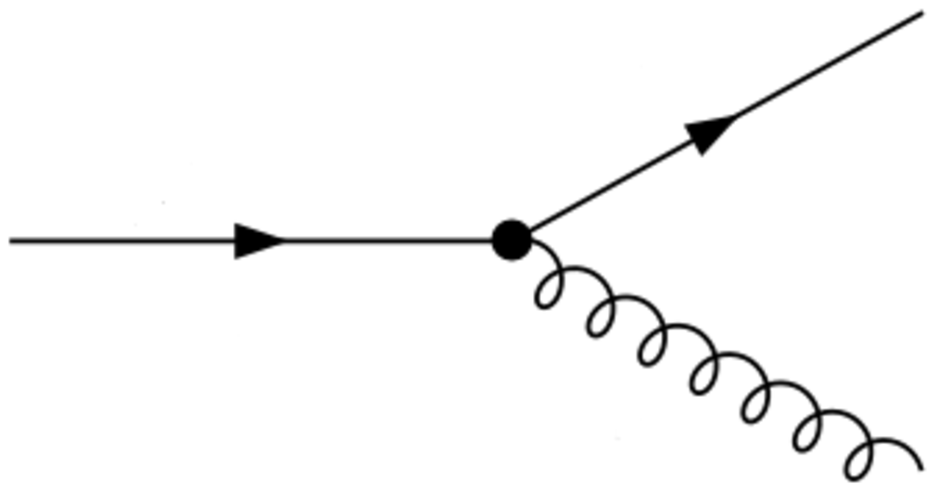}} \quad
 \subfloat[$e \gamma \to e \gamma $ vertex]{\includegraphics[width=0.25\textwidth]{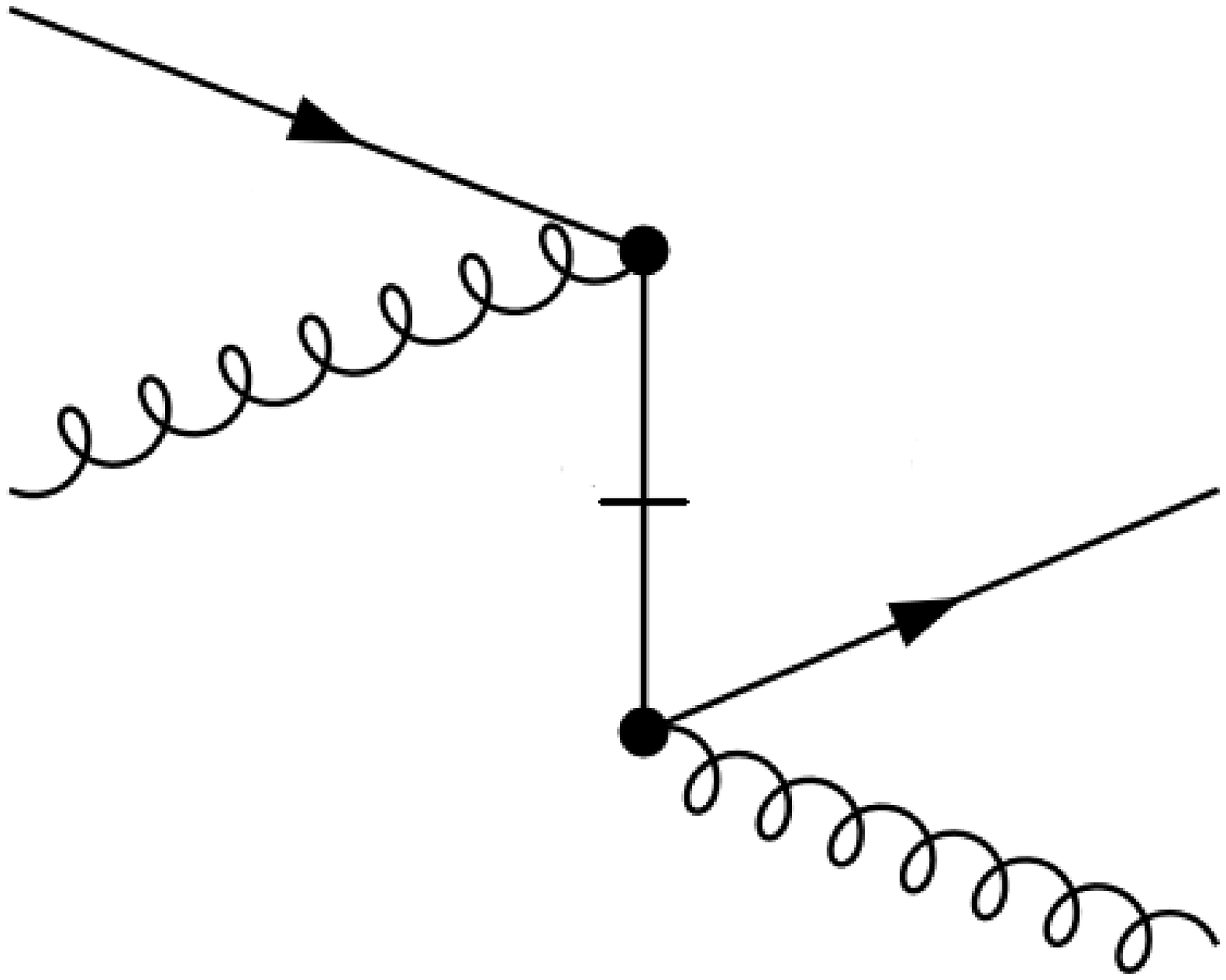}} \quad
 \subfloat[$ee \to ee$ vertex]{\includegraphics[width=0.25\textwidth]{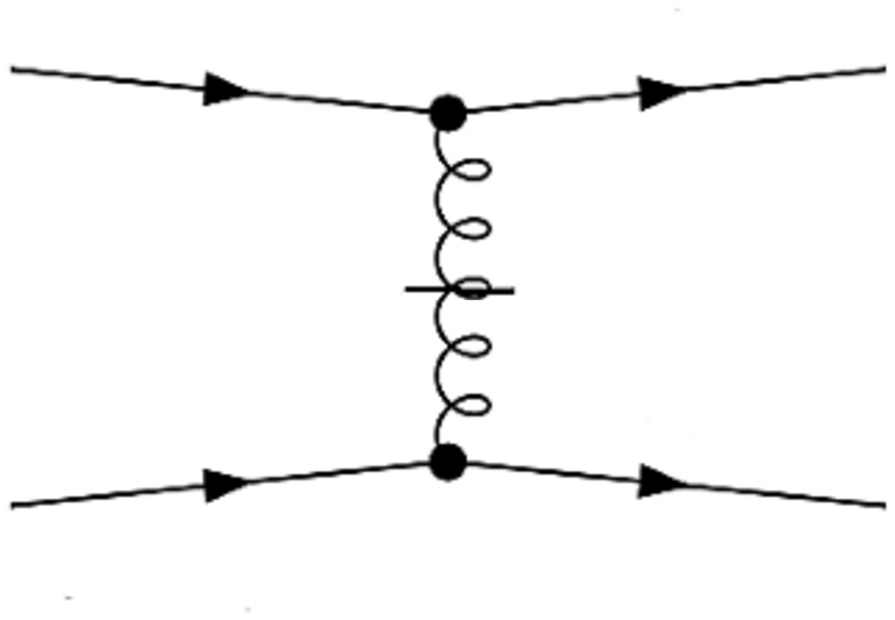}}
\caption{The relevant light-front QED vertices for positronium in the Fock sectors $\ket{e^- e^+}$ and $\ket{e^- e^+\gamma}$.}
\label{vertices}
\end{figure}

We adopt a light-cone gauge $A^+ = 0$ for QED. The light-front QED interactions are shown in Fig. \ref{vertices} which include the 
$e \to e \gamma$ vertex,
\begin{equation}
  P_+^{(e\to e\gamma)} =  \left. g \int \dd x_+\dd^2 x^\perp \; \bar{\psi}(x)\gamma_\mu \psi(x)  A^\mu(x) \right|_{x^+ = 0} \\
\end{equation}
and two instantaneous vertices,
 \begin{equation}
 \begin{split}
  P_+^{(e\gamma \to e\gamma)} &= \left. \half g^2 \int \dd x_+\dd^2 x^\perp \; \bar{\psi}(x) \gamma_\mu A^\mu(x) \frac{\gamma^+}{\imag \partial^+}\gamma_\nu A^\nu(x) \psi(x) \right|_{x^+ = 0}\\
  P_+^{(ee\to ee)} &= \left. \half g^2 \int \dd x_+\dd^2 x^\perp \; \bar{\psi}(x) \gamma^+ \psi(x) \frac{1}{(\imag \partial^+)^2} \bar{\psi}(x) \gamma^+ \psi(x) \right|_{x^+ = 0}.
 \end{split}
 \end{equation}

 In this model, we take the coupling constant $\alpha = \frac{g^2}{4 \pi} = 0.2$ and the basis energy scale $b = \sqrt{P^+\Omega} = 0.3 m_e$, where
 $m_e = 0.511 \;\mathrm{MeV}$ is the mass of the electron.  We adopt a regulator for the light-front small-$x$ singularity, 
 $\frac{1}{(x_1-x_2)^2} \to  \frac{1}{(x_1-x_2)^2 + \varepsilon}$ with $\varepsilon = 0.01$ \cite{Maris13, Karmanov10}.
\begin{figure}[h]
\centering
\includegraphics[width=0.9\textwidth]{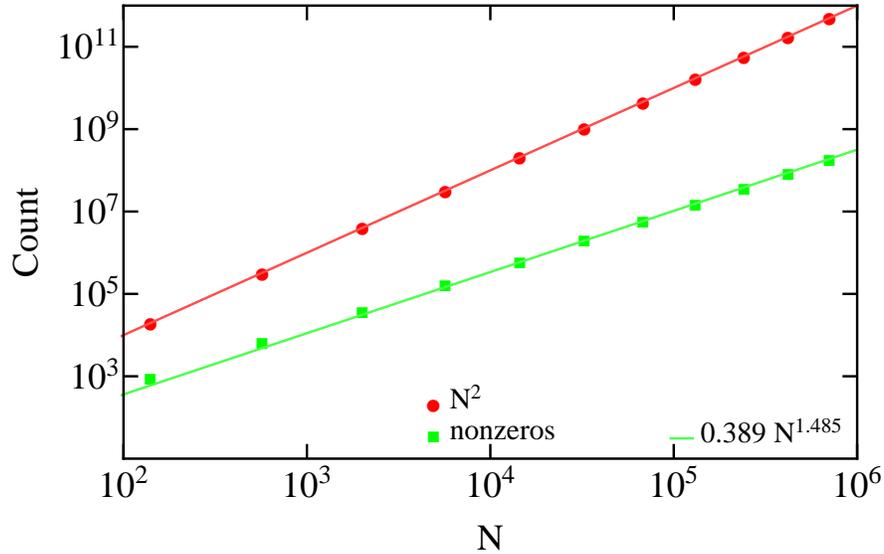} \quad
\caption{(Color Online) The sparsity of the light-cone Hamiltonian matrix $H_\textsc{lc}^{(\text{pos})}$. The horizontal axis is the dimensionality
 $N$ of the basis space. The square dots show the number of nonzero matrix elements. A power-law fitting suggests the number of nonzero elements is
 proportional to $\sim N^{1.485}$. }
\label{dimensionality}
\end{figure}
We then construct the light-cone Hamiltonian matrix $H_\textsc{lc}^{(\text{pos})}$. Fig. \ref{dimensionality} shows the number of nonzero matrix 
elements compared to the total number of matrix elements. The Hamiltonian matrix is a large sparse matrix. We diagonalize the matrix with the Lanczos method  
implemented by the ARPACK software \cite{Lehoucq97}, which is particularly suitable for solving large sparse 
matrix eigenvalue problems.
\begin{figure}[H]
 \centering
 \subfloat[]{\includegraphics[width=0.5 \textwidth]{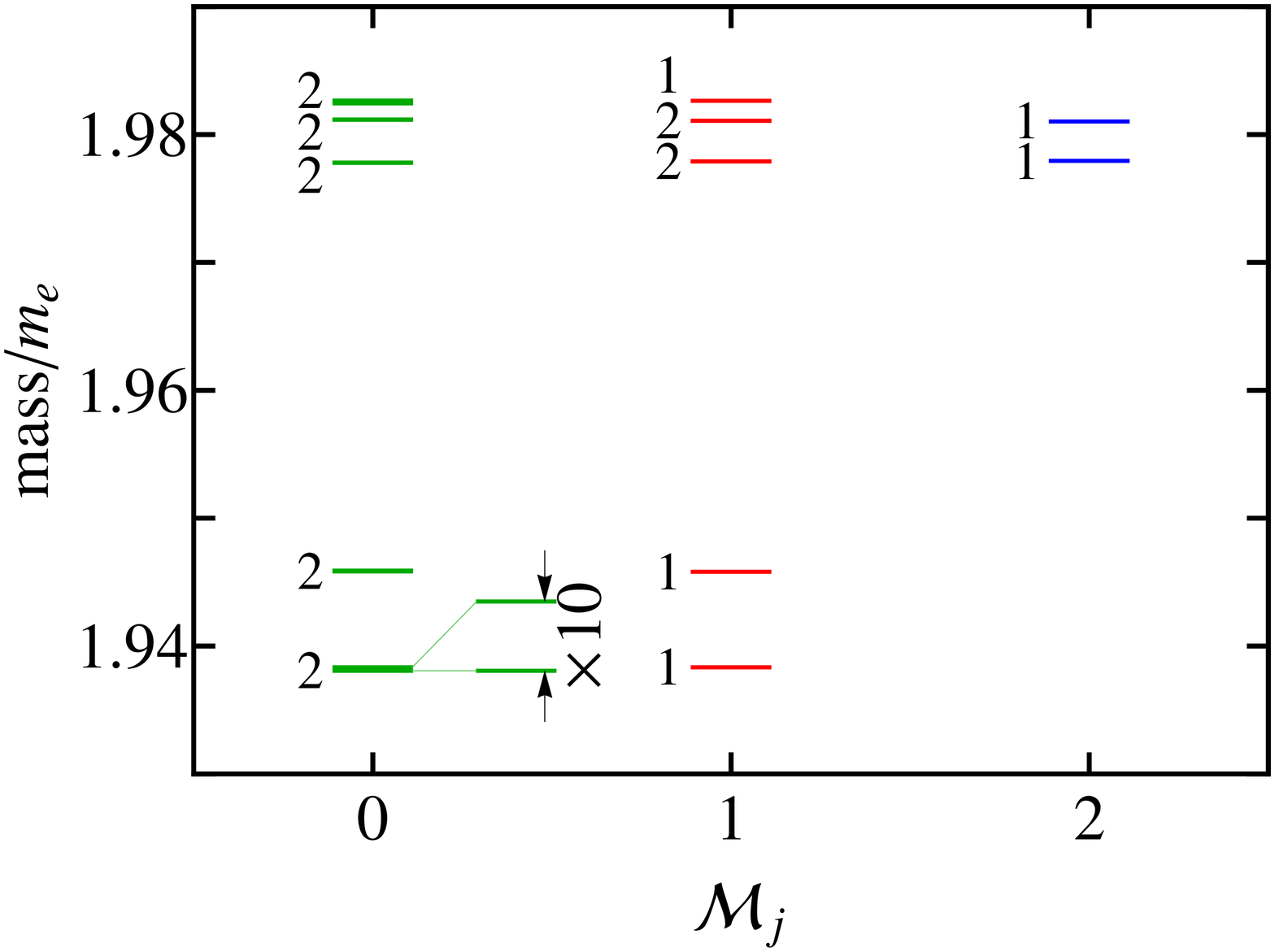}} 
 \subfloat[]{\includegraphics[width=0.5 \textwidth]{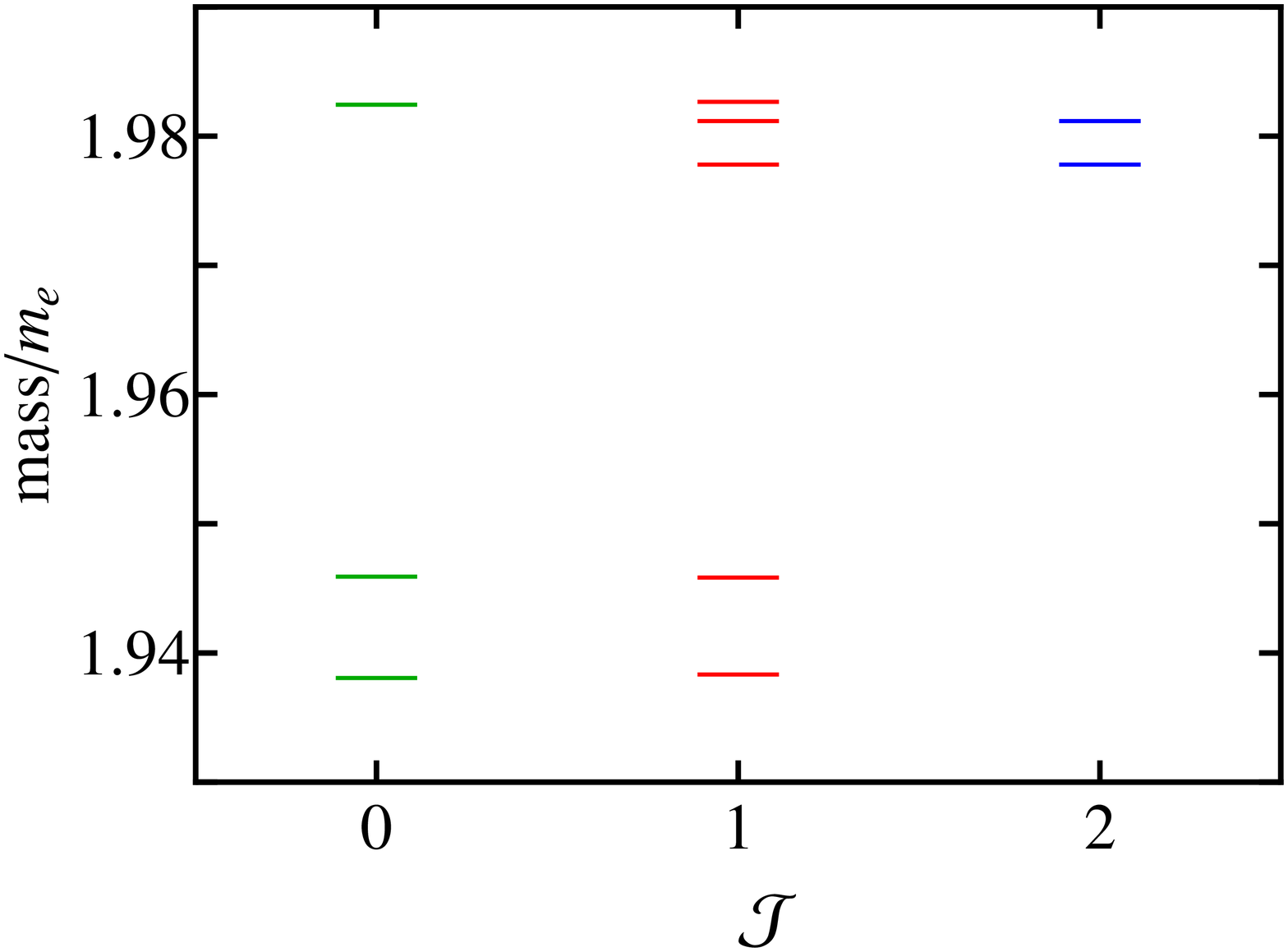}}
 \caption{(Color Online) The positronium spectrum for (a): spin projection $\mathcal{M}_j = 0,1,2$; (b): total spin $\J = 0, 1, 2, \mathcal{M}_j = 0$ identified from plot (a).
  For $\mathcal{M}_j = 3$, there is no state within the range of the plot. The numbers to the left of the bars in plot (a) label the multiplicity of the states 
around each bar. Plot (a) also shows the fine splitting between the lowest $\J = 0, \mathcal{M}_j = 0$ and $\J = 1, \mathcal{M}_j = 0$ states. The total spin $\J$ in plot (b) is 
obtained by counting the approximate degeneracy of states with different $\mathcal{M}_j$.}
\label{spectrum_b}
\end{figure}

We obtain the positronium spectrum directly from the diagonalization. Fig. \ref{spectrum_b}(a) shows the positronium bound-state spectrum for light-front
 spin projections $\mathcal{M}_j = 0, 1, 2$ from the numerical calculations with $\N = \K = 12, M_j = 0, 1, 2, 3; \lambda_\textsc{cm} = 5, \lambda_\textsc{b} = 0$. 
Some of the states in the figure are nearly degenerate. We have put a label $n$ to the left of each bar to indicate the existence of $n$ states around each bar in 
Fig. \ref{spectrum_b}(a). In the light-front TDA, the total spin $\J$ is only approximate. We can identify the total spin by counting the approximate 
degeneracy of particles with different $\mathcal{M}_j$ \cite{Brodsky98}. Fig. \ref{spectrum_b}(b) shows the states with $\mathcal{M}_j = 0$ identified by their total spin $\J$.
 The energy splitting between the singlet state ($\J = 0, \mathcal{M}_j = 0$) and the triplet state ($\J = 1, \mathcal{M}_j = 0$) is the fine splitting (Shown in Fig. \ref{spectrum_b}(a)).
 The non-relativistic quantum mechanics gives a fine splitting for the ground state $ E_{\text{triplet}} - E_{\text{singlet}} = \frac{1}{3} \alpha^4 m_e \simeq 5.33
 \times 10^{-4} m_e$ \cite{Griffiths08}. In our calculation, the fine splitting is $2.77 \times 10^{-4} m_e$, which is the correct order of magnitude and is 
reasonable given that we have a relativistic treatment (important for strong coupling) and our implementation of regulators.

\section{Conclusions and outlook}

 We have introduced the Basis Light-Front Quantization (BLFQ) approach to the QCD bound-state problem and analyzed the symmetries of the
 finite basis space. BLFQ converts the field theory problem into a form where we can take advantage of the recent advances in quantum
 many-body problems and the state-of-the-art methods developed for large sparse matrix eigenvalue problem. The compatible operators
 provide a means to identify states. We have shown the exact factorization of the center-of-mass motion in the finite basis space.
 They also allow us to reduce the dimensionality of the problem dramatically for a given level of accuracy, as we have demonstrated in
 the positronium model. BLFQ has retained the kinematic symmetries of the underlying Hamiltonian. We also introduced a generalized 2D
 Talmi-Moshinsky transformation to relate internal motions to a fixed coordinate system. Finally we have applied BLFQ to QED and obtained the 
positronium bound-state spectrum.

 To obtain physical results, one must perform renormalization. There are three types of divergences existing in the light-front quantized
field theory: the ultraviolet divergence, the infrared divergence and the spurious small-$x$ divergence. Various schemes have been developed
 to address these issues (for example \cite{Wiecki13, Perry07, Glazek95, Karmanov10}).

 As a non-perturbative \textit{ab inito} Hamiltonian approach to QFT, BLFQ requires major computational efforts. Thanks to the rapid advances
 in supercomputing, we foresee promising applications of BLFQ to understanding the structure of QCD bound states.

\section{Acknowledgement}
The authors thank S. J. Brodsky, H. Honkanen and D. Chakrabarti, V. Karmanov, A. El-Hady, C. Yang and E. G. Ng for fruitful discussions.
 This work was supported in part by the Department of Energy under Grant Nos. DE-FG02-87ER40371 and DESC0008485 (SciDAC-3/NUCLEI) and by
the National Science Foundation under Grant No. PHY-0904782. A portion of the computational resources were provided by the National Energy Research 
Scientific Computing Center (NERSC), which is supported by the DOE Office of Science.


\section*{Appendix A: Talmi-Moshinsky transformation}

Consider the exponential generating function of the 2D HO wavefunctions $\Psi_n^m(\bm{q})$,
\begin{equation}\label{HO generating function}
   e^{-\half\bm{q}^2 + 2 \bm{q}\cdot\bm{z} - \bm{z}^2}
= \sum_{n=0}^\infty \sum_{m=-\infty}^{\infty} \frac{(-1)^n}{\sqrt{4\pi(n+|m|)!n!}} \Psi_n^m(\bm{q}) e^{-\imag m \theta} z^{2n+|m|},
\end{equation}
where, $\bm{q} = \frac{\bm{p}}{\sqrt{x}}, z = |\bm{z}|/\sqrt{P^+\Omega}, \theta = \arg \bm{z}$. If we define,
\begin{equation}
\begin{split}
 \bm{Q} =  \bm{q_1} \cos \xi  + \bm{q_2} \sin \xi , &\quad
 \bm{q} =  \bm{q_1} \sin \xi  - \bm{q_2} \cos \xi , \\
 \bm{Z} =  \bm{z_1} \cos \xi  + \bm{z_2} \sin \xi , &\quad
 \bm{z} =  \bm{z_1} \sin \xi  - \bm{z_2} \cos \xi ,
\end{split}
\end{equation}
where $\xi = \arctan\sqrt\frac{x_2}{x_1}$, $\bm{Q} = \frac{\bm{p}_1+\bm{p}_2}{\sqrt{x_1+x_2}}, \bm{q} = \frac{x_2\bm{p}_1-x_1\bm{p}_2}{\sqrt{x_1x_2(x_1+x_2)}}$,
 then there exists an identity,
\begin{multline}
   ( -\half \bm{q}^2_1 + 2 \bm{q}_1\cdot \bm{z}_1 - \bm{z}_1^2) +  ( -\half \bm{q}^2_2 + 2 \bm{q}_2\cdot \bm{z}_2 - \bm{z}_2^2 ) = \\
 ( -\half \bm{Q}^2 + 2 \bm{Q} \cdot \bm{Z} - \bm{Z}^2 ) + ( - \half \bm{q}^2 + 2 \bm{q} \cdot \bm{z} - \bm{z}^2 ).
\end{multline}
Thanks to Eq.(\ref{HO generating function}),
\begin{multline}
 \sum_{n_1,m_1,n_2,m_2} \frac{(-1)^{n_1+n_2} \Psi_{n_1}^{m_1}(\bm{q_1}) \Psi_{n_2}^{m_2} (\bm{q_2})}{4\pi \sqrt{(n_1+|m_1|)!n_1! (n_2+|m_2|)!n_2!}}
 e^{-\imag m_1\theta_1-\imag m_2 \theta_2} z_1^{2n_1+|m_1|} z_2^{2n_2+|m_2|}	=\\
 \sum_{N,M,n,m} \frac{(-1)^{N+n}\Psi_N^M(\bm{Q})\Psi_n^m(\bm{q})}{4\pi\sqrt{(N+|M|)!N! (n+|m|)!n!}}
 e^{-\imag M \Theta - \imag m \theta} Z^{2N+|M|} z^{2n+|m|},
\end{multline}
where, $z_i = |\bm{z}_i|, Z = |\bm{Z}|, z = |\bm{z}|, \theta_i = \arg\bm{z}_i, \Theta = \arg\bm{Z}, \theta=\arg\bm{z}$.
We can express $Z, z, e^{\imag \Theta}, e^{\imag \theta}$ in terms of $z_1, z_2, e^{\imag \theta_1}, e^{\theta_2}$
and identify the corresponding coefficients.
\begin{equation*}
 \Psi_{n_1}^{m_1}(\bm{q}_1) \Psi_{n_2}^{m_2}(\bm{q}_2) \equiv \sum_{NMnm} \left. \left({NMnm}|{n_1m_1n_2m_2}\right)\right|_\xi
\delta_{\varepsilon_1+\varepsilon_2,E+\epsilon} \delta_{m_1+m_2,M+m} \Psi_N^M(\bm{Q})\Psi_n^m(\bm{q}),
\end{equation*}
where $E = 2N+|M|+1, \epsilon = 2n+|m|+1, \varepsilon_i = 2n_i+|m_i|+1$ and the coefficients are,
\begin{multline}
  \left.\left({N_1 M_1 N_2 M_2}|{n_1m_1n_2m_2}\right)\right|_\xi =  (-1)^{N_1-n_1-n_2+\half(|M_2|-M_2)}
(\sin\xi)^{2n_2+|m_2|} (\cos\xi)^{2n_1+|m_1|} \\
 \sqrt{\frac{(n_1+|m_1|)!(n_2+|m_2|)!n_1!n_2!}{(N_1+|M_1|)!(N_2+|M_2|)!N_1!N_2!}} \cdot
  \sum_{\gamma_1=0}^{v_1} \sum_{\gamma_2=0}^{v_2}	\sum_{\beta_1=0}^{\gamma_1} \sum_{\beta_2=0}^{\gamma_2} \sum_{\beta_3=0}^{V_2} \sum_{\beta=0}^{M_2}
   (-\tan\xi)^{\beta_1-\beta_2+\beta+M_2}  \\ \cdot {M_1 \choose \chi } \cdot {M_2 \choose \beta}
\cdot  {V_2 \choose \beta_1, \beta_2, \beta_3, \beta_4}
 \cdot { v_1+v_2-V_2 \choose \gamma_1-\beta_1, \gamma_2-\beta_2, v_1-\gamma_1-\beta_3, \beta_5},
\end{multline}
where $v_i = n_i + \half(m_i - |m_i|), V_i = N_i+\half(M_i-|M_i|), i=1,2$, $\chi = v_1-v_2+m_1+\gamma_2-\gamma_1-\beta$,  $\xi = \arctan \sqrt{\frac{x_2}{x_1}}$.
The multinomial coefficients ${n \choose m_1,m_2,\cdots, m_k} = \frac{n!}{m_1!m_2!\cdots m_k!}$ satisfy $m_1+m_2+\cdots m_k = n$, $0 \le m_i \le n, i=1,2,\cdots k$.
So, $\beta_4 = V_2 - \beta_1 - \beta_2 - \beta_3, \beta_5 = v_2 - \gamma_2 - \beta_4$.
The generalized binomial coefficients satisfy ${n \choose m} = \frac{n(n-1)\cdots(n-m+1)}{m!}, m \ge 0$ and ${n \choose 0} = 1$.


\end{document}